\documentclass[epjCONF,columns]{svjour} 
\pdfoutput=1
\usepackage{graphics}
\usepackage[varg]{txfonts} 
\usepackage[latin1]{inputenc}
\usepackage{cancel}
\usepackage{graphicx}

\session-title{Hadron Collider Physics Symposium - 2011}
\begin{document}
\title{Searches for Low Mass Higgs Boson at the Tevatron}
\author{Federico Sforza\thanks{\email{federico.sforza@pi.infn.it}} for the CDF and D0 Collaborations}
\institute{University \& INFN Pisa}
\abstract{
  We present the result of the searches for a low mass Standard Model Higgs 
  boson performed at the Tevatron $p\bar{p}$ collider ($\sqrt{s} =1.96$~TeV) by the CDF and 
  D0 experiments with an integrated luminosity of up to $8.5$~fb$^{-1}$.
  Individual searches are discussed and classified  according to their sensitivity.
  Primary channels rely on the associate production with a vector boson ($WH$ or $ZH$) 
  and the $H\to b\bar{b}$ decay channel (favored for $M_H\lesssim 135$~GeV$/c^2$). Event selection is based 
  on the leptonic decay of the vector boson and the identification of $b-$hadron enriched jets. Each 
  individual channel is sensitive, for $M_H=115$~GeV/$c^2$, to less than $5$ times the SM expected cross 
  section and the most sensitive channels can exclude a production cross section of $2.3\times\sigma_H^{SM}$
  Secondary channels rely on a variety of final states. 
  Although they are from $2$ to $5$ times less sensitive than any primary channel, they contribute to the 
  Tevatron combination and, in some cases, they pose strong constrains on exotic Higgs boson models.
} 
\maketitle
\section{Introduction}\label{intro}

The spontaneous symmetry breaking mechanism~\cite{higgs_b} offers a possible explanation
for $W$ and $Z$ boson mass within the Standard Model~\cite{SM_b} of particle physics.
A new scalar particle, the Higgs boson, is predicted but direct experimental confirmation is
missing. 

In this paper we summarize the direct searches performed at the Tevatron $p\bar{p}$
collider ($\sqrt{s}=1.96$~TeV) by the CDF and D0 experiments with the summer 2011 dataset,
corresponding to an integrated luminosity of up to $8.5$~fb$^{-1}$. Analyses are optimized
for the low range of allowed Higgs boson masses: $100\lesssim M_H\lesssim 135$~GeV$/c^2$.
This range is favored by indirect constraints coming from the measurement of
other SM parameters~\cite{gfitter_b}.

\section{Low Mass Higgs Analyses}
\begin{figure}
  \begin{center}
    \resizebox{0.85\columnwidth}{!}{%
      \includegraphics{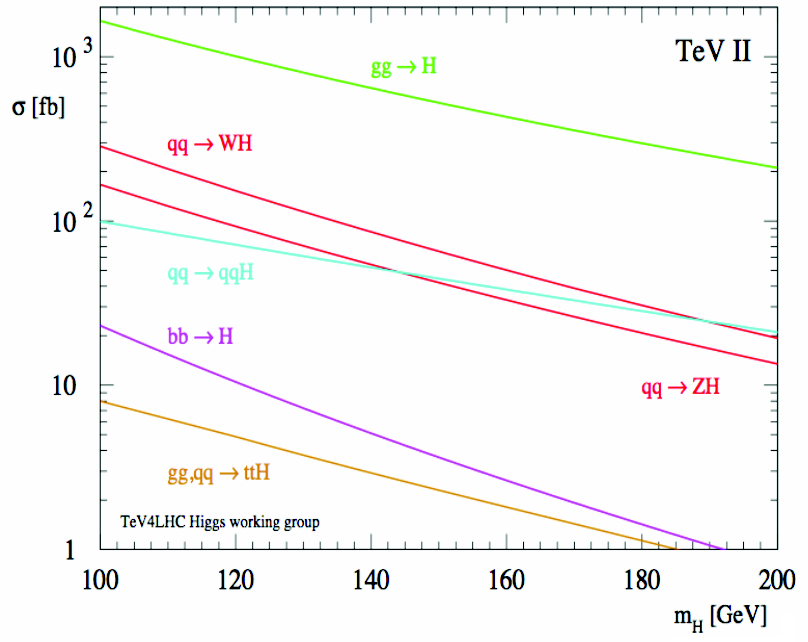} }
  \end{center}
\caption{Cross section of the different processes contributing to the SM Higgs boson 
  production at the Tevatron $p\bar{p}$ collider ($\sqrt{s}=1.96$~TeV) as a function 
  of $M_H$.}
\label{fig:higgsCX} 
\end{figure}

\begin{figure}
  \begin{center}
    \resizebox{0.75\columnwidth}{!}{%
      \includegraphics{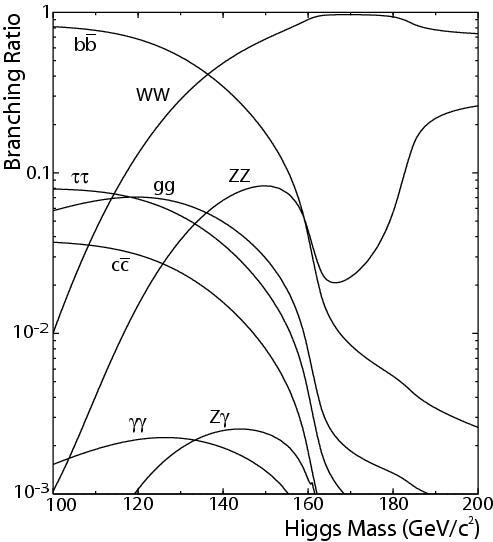} }
  \end{center}
\caption{SM Higgs boson branching fractions depending on the considered $M_H$.}
\label{fig:higgsBR} 
\end{figure}

\begin{sloppypar} 
The hadron collider environment is experimentally complex
because of the overwhelming background of multi-jet events hiding rare processes such as 
Higgs boson production. We need a distinct event 
signature to increase the signal over background ratio and thus the sensitivity.
The individual analyses can be classified in two classes on the basis of the final states 
and the expected sensitivity:
\begin{description}
\item[Primary Channels:] they identify the most sensitive analyses and
  they all share common characteristics. The Higgs boson is produced in association with a 
  $W$ or a $Z$ bosons (see Figure~\ref{fig:higgsCX} for predicted cross sections) and the 
  leptonic decay of the vector boson is used for the online 
  and offline event selection. The Higgs candidates are selected in the $b\bar{b}$
  final state, as Figure~\ref{fig:higgsBR} shows, this is the
  favored final state (for $M_H\lesssim 135$~GeV$/c^2$) because of the Yukawa coupling
  of the SM Higgs boson to the fermions~\cite{higgs_b}.
  More details about the analysis techniques are given (Section~\ref{subSec:tech})
  because of the relevant impact of these analyses.
\item[Secondary Channels:] the Higgs is produced via gluon fusion or $t\bar{t}$ 
  associate production (see Figure~\ref{fig:higgsCX}). Each analysis is optimized for a 
  different final state appearing in 
  Figure~\ref{fig:higgsBR}: $\gamma\gamma$, $\tau\tau$, high $b-$jets multiplicity. 
  Each secondary channel analysis is from 2 to 5 times less sensitive than any primary channel
  but their contribution is not negligible when considered all together. Furthermore, many
  non SM Higgs boson scenarios predict a production rate which increase in these final states.
\end{description}
\end{sloppypar}

\section{Primary Search Channels}
\label{sec:primCh}

All primary channels analyses look for the $H\to b\bar{b}$ decay when produced in association 
with a $W$ or $Z$ boson that undergo a leptonic decay. They are classified by the 
following signatures:
\begin{eqnarray}
  ZH\to\ell\ell+b\bar{b}\mathrm{,} & WH\to\ell\cancel{\ell} +b\bar{b}\mathrm{,} &  W/Z H\to \cancel{\ell}\cancel{\ell} +b\bar{b}
\end{eqnarray}
where $\ell$ is an electron or a muon and $\cancel{\ell}$ indicates a neutrino or a lepton 
which has not been identified (e.g. $W\to \tau\nu$ events) appearing as an imbalance in the 
total transverse energy ($\cancel{E}_T$). 
Both CDF and D0 collaborations performed analyses requiring two leptons~\cite{cdfll_b,d0ll_b}, one lepton
plus $\cancel{E}_T$~\cite{cdflv_b,d0lv_b} and $\cancel{E}_T$ only~\cite{cdfvv_b,d0vv_b}. 
Although higher lepton multiplicity usually corresponds to a cleaner 
signature, the channels with one or no lepton identified are slightly more sensitive because of the 
higher selection efficiency and $WH$ production cross section (see Figure~\ref{fig:higgsCX}).

Background composition is another common feature of the three primary channels both in CDF and D0 
analyses, we can divide it in three categories: the larger is the $W/Z$ production in association with
light and heavy flavor jets, after selection this irreducible background can be from $10^2$ to $10^4$ 
larger than the expected signal. The second background is due to multi-jet events faking the $\cancel{E}_T$ and
the lepton identification. The contribution of real physics processes and detector effects makes this
background particularly difficult to model so the contamination should be reduced as much as possible 
at selection level. The last category is composed by smaller electroweak processes like top-quark or
diboson production. 

Because of the small Higgs production cross section, the final challenge is the 
maximization of the acceptance while keeping the backgrounds under control. 
Single-top observation~\cite{cdfsingle_top_b,d0single_top_b} and 
diboson evidence in the heavy flavor final states~\cite{cdfdiboson_b,d0diboson_b} 
already demonstrate that the 
Tevatron experiments can probe sub-picobarn cross sections in these channels.

\subsection{Analysis Techniques}
\label{subSec:tech}

The analysis process of the primary channels can be divided into four stages: online selection, 
offline lepton selection, application of b-tagging algorithm and evaluation of the final
discriminant. In the CDF and D0 analyses, each of these stages have been highly optimized,
often thanks to the use of multivariate techniques: Neural Networks (NN), 
Boosted Decision Trees (BDT) or Support Vector Machines (SVM)~\cite{Multivariates_b}.
Machine learning algorithms are powerful regression or classification tools
as they can exploit the non-linear correlations between several input variables.
The reliability of their results is ensured by
checking the input and output distribution between the training samples and
the data in various control regions.

The first stage of the analysis 
process is the online selection. Collision events are collected and recorded via 
dedicated trigger paths that meet specific physics goals (e.g. high-$P_T$ lepton 
identification). Combining multiple paths
maximizes the acceptance, however the trigger efficiency must be parametrized properly
on Monte Carlo events. Especially in multiple-objects trigger paths (e.g. $\cancel{E}_T$ plus jets)
the efficiency function may depend on many variables.
At CDF, for the first time~\cite{cdfll_b,cdfvv_b} we used a NN to model the probability 
distribution of events selected by a large set of triggers.

The next stage is the offline event selection. We gained acceptance with
the introduction of more lepton categories (track only reconstruction, 
likelihood and NN identification, etc.) and relaxing the
cuts on jets and $\cancel{E}_T$ selection~\cite{cdfvv_b,d0vv_b}. This increased the multi-jet
background but multivariate techniques proved to be extremely effective to reduce it
and keep it under control~\cite{cdflv_b,cdfvv_b,d0vv_b}.

The last selection stage is the application of b-tagging algorithms to select 
jets enriched in heavy flavors. The identification of $b\bar{b}$ events can 
reduce the $W/Z$ plus jets (of generic flavor) background by a factor of 100 although 
at the cost of a significant inefficiency on signal ($\epsilon_{b-tag}\lesssim 50$\%).
CDF and D0 collaborations undertook a strong effort on b-tagging strategy optimization.
The CDF analyses combine the response of three different b-taggers: the 
SECVTX~\cite{secVtx_b} algorithm identifies displaced secondary vertexes, the JETPROB~\cite{jetProb_b}
algorithm exploits the impact parameter of the tracks and, finally, the information
of a neural-network-based tagger can also be used.
The D0 analyses deploy a BDT~\cite{d0tags_b} algorithm that includes
information relating to the lifetime of the hadrons in the jet. The result is a continuous
variable discriminating between $b$ and light jets.

After the selection is complete, we remain with a $W/Z$ plus heavy flavor jets sample
that could contain the Higgs signal. The invariant mass of the di-jet system is, by definition, 
the variable that distinguish a resonance over a non-resonant background produced by
QCD interactions, so, in this case, the final sensitivity is limited by the jet energy resolution.
Also other variables with smaller separation power exists, CDF combines them using Neural Networks 
(up to 7 variables are used) while D0 analyses use BDT (with up to 32 variables).
The use of multivariate techniques improves the final sensitivity up to 20\% over the simple invariant
mass approach.

\subsection{$H\to b\bar{b}$ Sensitivity}
\label{subSec:hbbRes}

The final expected and observed sensitivity of the individual primary search channels for 
different $M_H$ are summarized in 
Table~\ref{tab:cdfVH_lim} for CDF and Table~\ref{tab:d0VH_lim} for D0. The most
sensitive analysis can exclude at $95$\% C.L. the presence of a Higgs boson of 
$M_H=115$~GeV$/c^2$ produced with a cross section of 2.3 times the one predicted by the SM.

\begin{figure}
\resizebox{0.98\columnwidth}{!}{%
  \includegraphics{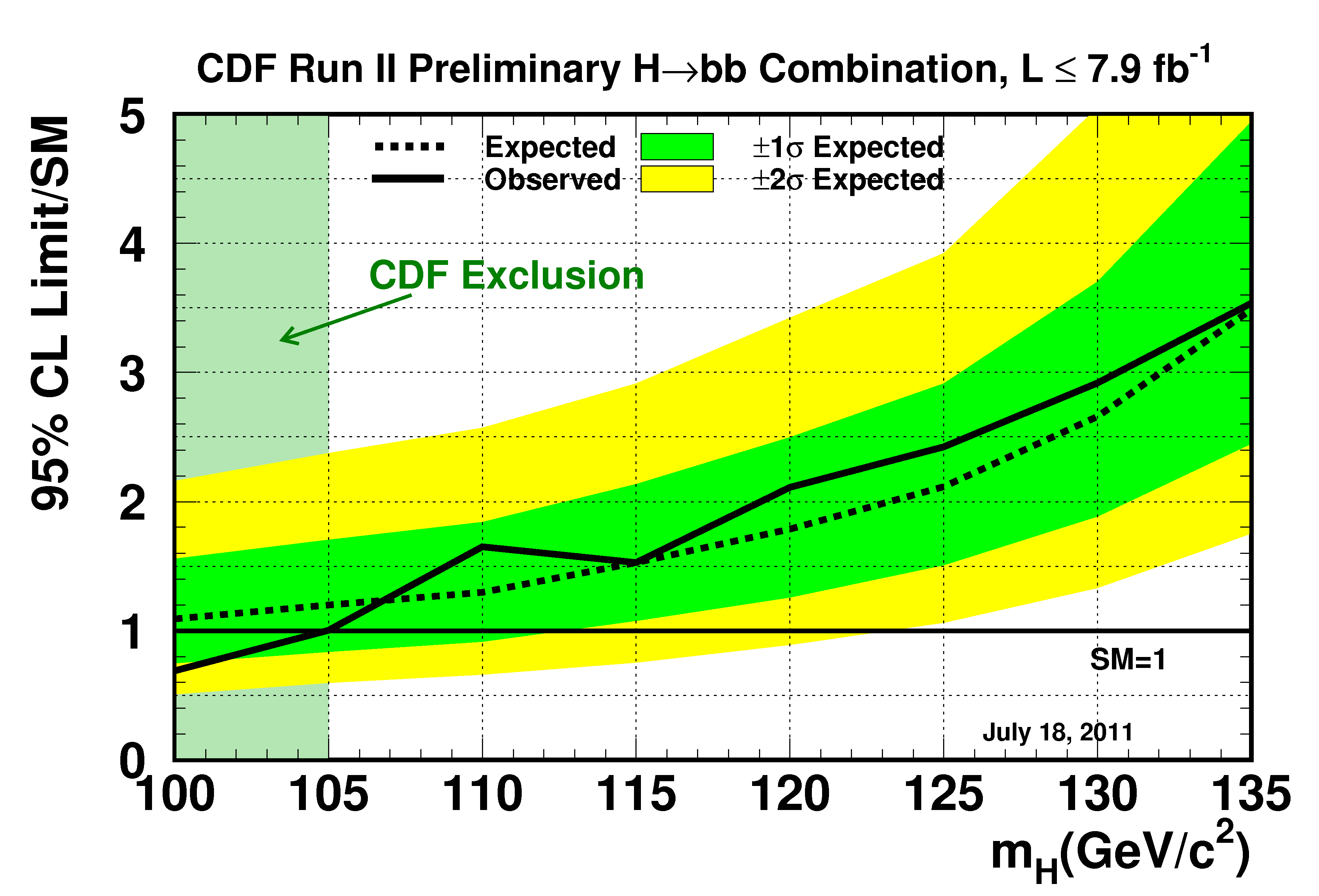} }
\caption{ble}
\caption{Observed and expected $95$\% C.L. measured by the CDF experiment
  using a luminosity up to $7.8$~fb$^{-1}$ for different SM Higgs boson 
  masses and with a final state $H\to b\bar{b}$. The limit is obtained 
  combining all the CDF primary search channels as they share the $H\to b\bar{b}$
  signature.}

\label{fig:cdfhbb} 
\end{figure}

Individually none of these analyses reaches the SM sensitivity for any analyzed 
$M_H$, however each experiment can combine the three channels to obtain
powerful constraints on the $H\to b\bar{b}$ production and decay. The combination of different 
channels, across the same experiment, provides also an advantage in the evaluation of the 
correlated systematic uncertainties. For example effects like Jet Energy Scale (JES) uncertainty 
or b-tag efficiency measurement on MC are shared across all the channels and we fit for their  
best value~\cite{cdflowmass_b}, in this way a higher statistical sample poses a stronger constraint on
these systematics than each channel by itself.
Figure~\ref{fig:cdfhbb} shows that the CDF experiment by itself excludes, at $95$\% C.L., the presence 
of a Higgs boson for $M_H<105$~GeV$/c^2$, giving an independent confirmation of the LEP~\cite{lep_b}
exclusion in the same region. The exclusion limit can
be extended further combining the results from both the CDF and D0 experiments~\cite{tevComb_b}.

\begin{table}
\caption{Observed and expected $95$\% C.L. measured by the CDF experiment
  using a luminosity up to $7.8$~fb$^{-1}$ for different SM Higgs boson masses in the
  {\em primary search channels} ($\ell\ell + b\bar{b}$, $\ell\cancel{\ell}+b\bar{b}$, $\cancel{\ell}\cancel{\ell}+b\bar{b}$).}
\label{tab:cdfVH_lim}       
\begin{tabular}{llllllll}
\hline\noalign{\smallskip}
$M_H$ (GeV$/c^2$) & 100 & 105 & 110& 115 &120 &125& 130 \\
\noalign{\smallskip}\hline\noalign{\smallskip}
\multicolumn{8}{l}{$Z H\to \ell\ell +b\bar{b}$}\\
Exp.& 2.7& 3.1& 3.4& 3.9& 4.7& 5.5 & 7.0\\
Obs.& 2.8& 3.3& 4.4& 4.8& 5.4& 4.9 & 6.6\\
\noalign{\smallskip}\hline\noalign{\smallskip}
\multicolumn{8}{l}{$W H\to \ell\cancel{\ell} +b\bar{b}$}\\
Exp.& 1.8 & 2.0& 2.2& 2.6& 3.1& 3.7& 4.8\\
Obs.& 1.1 & 2.1& 2.8& 2.7& 3.4& 4.4& 6.1\\

\noalign{\smallskip}\hline\noalign{\smallskip}
\multicolumn{8}{l}{$V H\to \cancel{\ell}\cancel{\ell} +b\bar{b}$}\\
Exp. &2.3 & 2.4& 2.6& 2.9& 3.4& 4.0& 4.9\\ 
Obs. &1.8 & 1.8& 2.2& 2.3& 3.3& 5.4& 5.0\\

\noalign{\smallskip}\hline
\end{tabular}
\end{table}

\begin{table}
  \caption{Observed and expected $95$\% C.L. measured by the D0 experiment 
    using a luminosity up to $8.5$~fb$^{-1}$ for different SM Higgs boson masses 
    in the{\em primary search channels} ($\ell\ell+b\bar{b}$, $\ell\cancel{\ell}+b\bar{b}$, $\cancel{\ell}\cancel{\ell}+b\bar{b}$).}
\label{tab:d0VH_lim}       
\begin{tabular}{llllllll}
\hline\noalign{\smallskip}
$M_H$ (GeV$/c^2$) & 100 & 105 & 110& 115 &120 &125& 130 \\
\noalign{\smallskip}\hline\noalign{\smallskip}
\multicolumn{8}{l}{$Z H\to \ell\ell +b\bar{b}$}\\
Exp. &3.4& 3.7& 4.2& 4.8& 5.3& 6.5& 8.4\\
Obs. &2.5& 2.6& 3.1& 4.9& 6.4& 8.9& 9.9\\

\noalign{\smallskip}\hline\noalign{\smallskip}
\multicolumn{8}{l}{$W H\to \ell\nu +b\bar{b}$}\\
Exp. &2.4 &2.6& 3.0& 3.5& 4.3& 5.4& 7.0\\
Obs. &2.6 &2.9& 4.1& 4.6& 5.8& 6.8& 8.2\\

\noalign{\smallskip}\hline\noalign{\smallskip}
\multicolumn{8}{l}{$V H\to \cancel{\ell}\cancel{\ell} +b\bar{b}$}\\
Exp. &2.8& 2.9& 3.1& 4.0& 4.5& 5.4& 6.9\\
Obs. &2.6& 2.4& 2.4& 3.2& 3.9& 5.0& 7.5\\

\noalign{\smallskip}\hline
\end{tabular}
\end{table}

\section{Secondary Search Channels}
\label{sec:secChan}

The primary channels described in the previous sections play a major role in the Higgs boson 
searches performed by the CDF and D0 collaborations, however there are a variety of final states
worth investigating. The most significant are the $H\to \gamma\gamma$~\cite{cdfhgg_b,d0hgg_b},
and $H\to \tau\tau$~\cite{cdflowmass_b,d0htt_b} decay channels and the 
$ttH\to l\nu b\bar{b}b\bar{b}$ associate production~\cite{cdftth_b}.
For example, even though the diphoton final state has a tiny branching fraction, 
it can still contribute significantly to the low mass Higgs boson searches 
due to better mass resolution and detector acceptance relative to b-quark final states:
for $M_H=115$~GeV$/c^2$, expected sensitivity of $11\times\sigma_{H}^{SM}$ are reached by D0 and $13\times\sigma_{H}^{SM}$ by
CDF.
Similar contribution comes from the $H\to \tau\tau$ channel (expected sensitivity
is $12.8\times\sigma_{H}^{SM}$ for D0 and $12.6\times\sigma_{H}^{SM}$ for CDF for $M_H=115$~GeV$/c^2$)
because the branching ratio of the SM Higgs boson to a $\tau$ pair is the second highest
(7.6\% at $M_H = 115$~GeV$/c^2$); the accurate knowledge of the $Z\to \tau\tau$ process
also helps the analysis of this channel.
In general each of the secondary channels reach a sensitivity on the order of
$12\times\sigma_{H}^{SM}$
and the composition of all of them contributes to the final search at the level of an 
additional primary channel~\cite{cdfcomb_b,d0comb_b}.

\begin{figure}
\resizebox{0.99\columnwidth}{!}{%
  \includegraphics{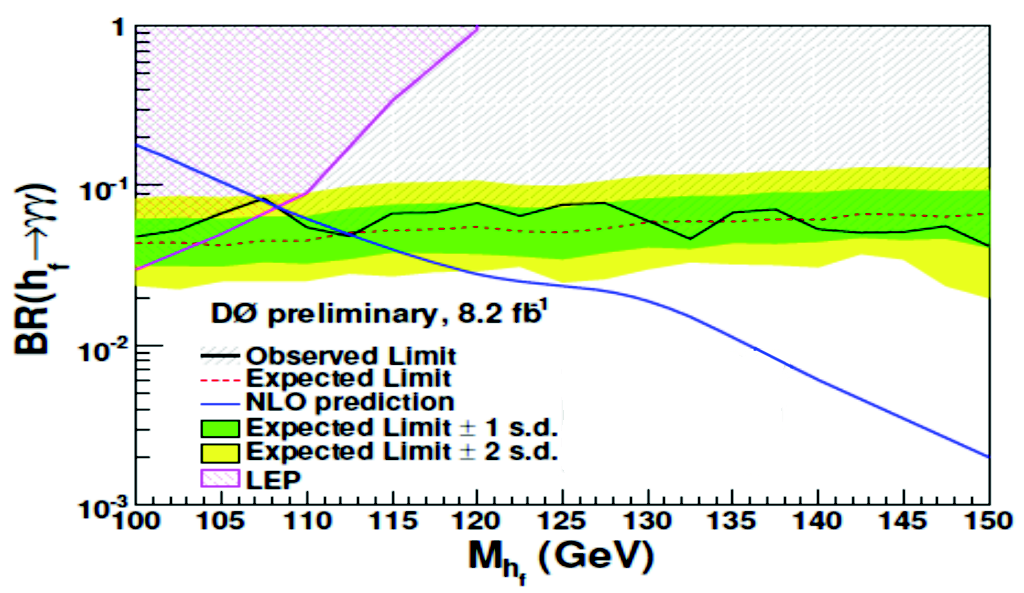} }
\caption{Observed and expected $95$\% C.L. measured by the D0 experiment
  using a luminosity of $8.2$~fb$^{-1}$ in the $H\to \gamma\gamma$ channel
  for a fermiophobic Higgs boson of different masses.}
\label{fig:d0_fpho} 
\end{figure}

Another reason to pursue these secondary channels is that non-SM theories  
may predict enhanced yield. For example the $H\to \gamma \gamma$ analyses
can be reinterpreted in the light of a fermiophobic Higgs boson theory where the 
couplings to the fermions are depressed~\cite{fpho_teo_b}. Figure~\ref{fig:d0_fpho} shows the
exclusion limits posed by the D0 collaboration for this particular model.

\section{Results and Future Prospects}
\label{sec:results}
\begin{sloppypar} 
The CDF and D0 experiments performed a variety of searches for a low mass Higgs boson.
The favored SM decay channel ($H\to b\bar{b}$) is analyzed in the associated
production modes, $WH$ and $ZH$, where the leptonic decay of the vector boson
allows an efficient online selection and offline reconstruction of the candidates. 
The expected sensitivity, for the best channel and $M_H=115$~GeV$/c^2$, 
reaches $2.6\times\sigma_H^{SM}$. The $H\to b\bar{b}$ channels have been combined
within each experiment to exclude at $95$\% C.L. Higgs boson production for $M_H<105$~GeV$/c^2$.

Also the results of a variety of less favorite search channels is analyzed: 
$H\to \gamma\gamma$, $H\to \tau\tau$, $ttH$ associate production.
They reach approx $12\times\sigma_H^{SM}$ sensitivity (for $M_H=115$~GeV$/c^2$).
Furthermore they can strongly constrain
exotic models, for example the fermiophobic Higgs model has been excluded (at $95$\% C.L.) up to
$M_H<109$~GeV$/c^2$~\cite{cdfhgg_b,d0hgg_b}.

\begin{figure}
\resizebox{0.99\columnwidth}{!}{%
  \includegraphics{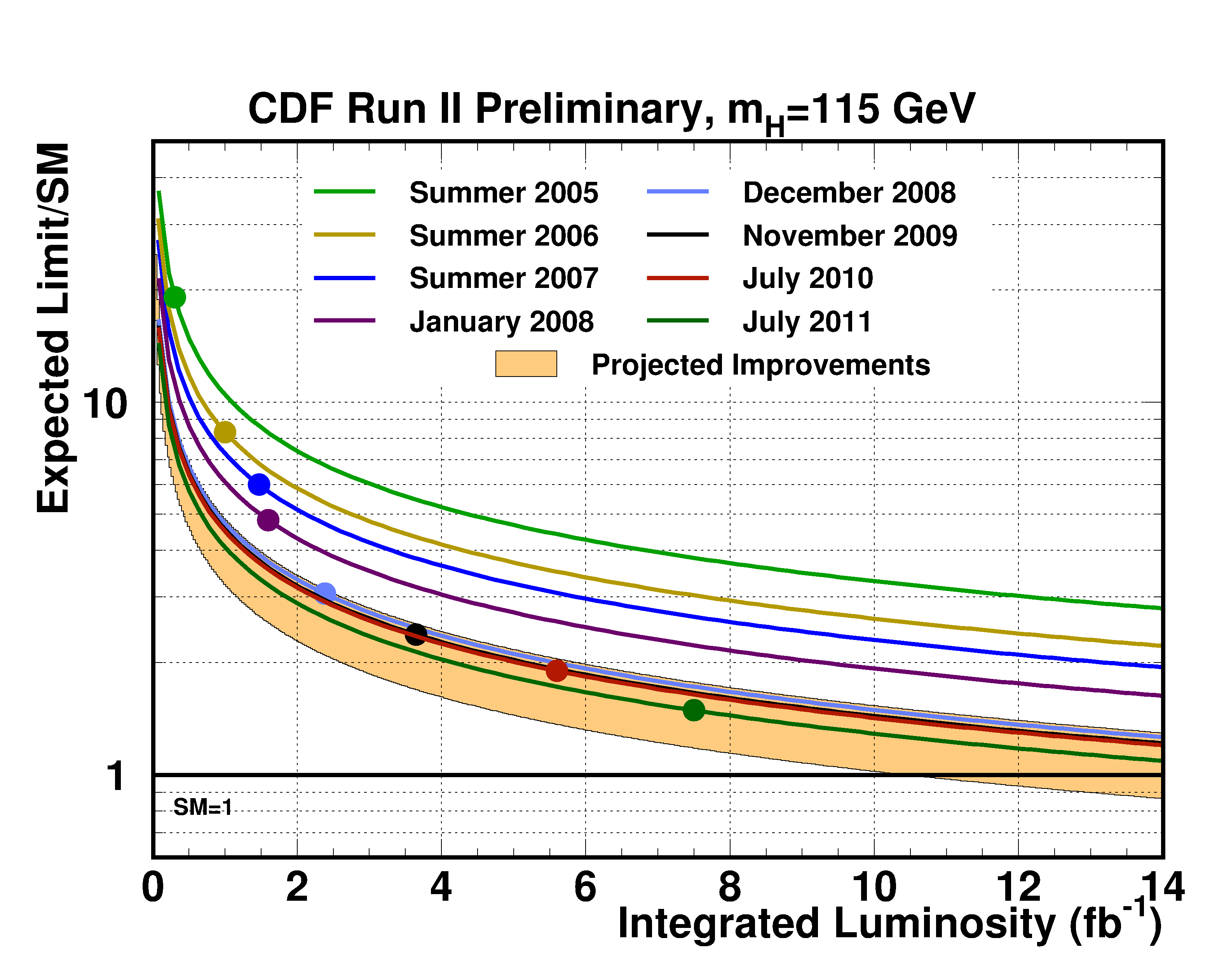} }
\caption{Plausible scenario of the analysis improvements that can be finalized
  by the CDF collaboration (similar results are expected from D0) 
  for the Winter 2012 conferences where the full dataset of $10$~fb$^{-1}$ will be analyzed.}
\label{fig:cdfimprov} 
\end{figure}

Figure~\ref{fig:cdfimprov} shows a plausible scenario of the analysis improvements that can be finalized
by the CDF collaboration (similar results are expected from D0) 
for the Winter 2012 conferences when the full dataset of $10$~fb$^{-1}$ will be analyzed. 
Thanks to the improvements planned by the CDF and D0 collaborations, we expect that the
Tevatron will reach the sensitivity needed to exclude at 95\% C.L. a SM Higgs boson across the
mass range $100\lesssim M_H\lesssim 135$~GeV$/c^2$~\cite{tevComb_b}.
\end{sloppypar}

\section*{Acknowledgments}

We would like to thank the organizers of the 2011 Hadron
Collider Physics Symposium for a wonderful conference
with excellent presentations and the CDF and D0 collaborations
for the results presented at this conference.

\end{document}